\documentclass[a4paper,usenatbib,useAMS]{mn2e}

\usepackage{graphicx}	
\usepackage{amsmath}	
\usepackage{amssymb}	
\usepackage{multicol}        
\usepackage{bm}		
\usepackage{pdflscape}	

\usepackage{amsfonts}
\usepackage{epsfig,graphics}
\usepackage{color}
\usepackage{subfigure}
\usepackage{caption}
\usepackage{tikz}
\usepackage{pgfplots}
\usepackage{amssymb}
\usepackage{graphicx}
\DeclareGraphicsExtensions{.pdf .png}

\newcommand{\m}[1]{\begin{math} {#1} \end{math}}

\usepackage[T1]{fontenc}
\usepackage{ae,aecompl}

\title[Calibrating Photo-z with Cross Correlations]{A Joint Analysis for Cosmology and Photometric Redshift Calibration Using Cross Correlations} 

\author[M. McLeod, F. B. Abdalla, S. Balan]{Michael McLeod$^1$\thanks{Email:
         michael.mcleod.13@ucl.ac.uk}, Sreekumar T. Balan$^1$, Filipe B. Abdalla$^{1,2}$
\\
$^1$ Department of Physics and Astronomy, University College London, Gower Place, London WC1E 6BT, U.K.\\
$^2$ Department of Physics and Electronics, Rhodes University, PO Box 94, Grahamstown, 6140, South Africa}

\pubyear{2016}

\begin{document}
\label{firstpage}
\pagerange{\pageref{firstpage}--\pageref{lastpage}}
\maketitle
%

\begin{abstract}
We present a method of calibrating the properties of photometric redshift bins as part of a larger Markov Chain Monte Carlo (MCMC) analysis for the inference of cosmological parameters. The redshift bins are characterised by their mean and variance, which are varied as free parameters and marginalised over when obtaining the cosmological parameters. We demonstrate that the likelihood function for cross-correlations in an angular power spectrum framework tightly constrains the properties of bins such that they may be well determined, reducing their influence on cosmological parameters and avoiding the bias from poorly estimated redshift distributions. We demonstrate that even with only three photometric and three spectroscopic bins, we can recover accurate estimates of the mean redshift of a bin to within \m{\Delta\mu \approx 3-4 \times10^{-3}} and the width of the bin to \m{\Delta\sigma \approx 1\times10^{-3}} for galaxies near \m{z = 1}. This indicates that we may be able to bring down the photometric redshift errors to a level which is in line with the requirements for the next generation of cosmological experiments. 
\end{abstract}




\section{Introduction} \label{intro}

Galaxy surveys have become in recent years an important source of data for cosmology, particularly for late time effects such as dark energy. Calculations of predicted statistical properties for a given cosmological model require the redshift distribution of observed galaxies to be known accurately. Spectroscopy has long been used to calculate accurate redshifts for objects, but this is a time intensive process requiring both detailed observation across the object's spectrum and careful analysis. In order to collect data for the vast numbers of galaxies required, current and future surveys are necessarily dependent on photometric redshifts for the majority of objects. 

When photometry is used, approximate redshifts are calculated from a small number of intensities measured in (typically around five) broad bands. Standard methods of inferring redshifts from photometric data are to use machine learning methods, such as artificial neural networks, or to fit template functions (see e.g. \citealt{photoz_review} for a review). These require that we have large training sets of galaxies for which we have spectroscopic redshifts. Additionally, the spectroscopic set must be representative of the full photometric set (in terms of both redshift range \emph{and} the nature of the objects contained within the sample) in order to reduce both the error and the bias in the derived relation (\citealt{ANNz2}, \citealt{photoz_compare}). Unfortunately, the spectroscopic sample is rarely as large as we would like, and is even less often fully representative of the redshift range we wish to look at. Spectroscopic samples tend to be dominated by bright objects which are easier to study; a lack of spectroscopic objects, particularly at the extremes of the redshift range, tends to lead to larger errors in the redshift distributions reconstructed from machine learning techniques. (Often the middle of the range is reconstructed comparatively well, while the outer regions suffer.) Other methods may be more successful at the lower or higher end of redshift, but few methods can be confidently used across the entire range (\citealt{photoz_compare}). In addition to this, the errors associated with any such reconstructions are large, and those who have attempted to reconstruct redshifts from photometric data will be familiar with the significant scatter around the spectroscopic redshifts (e.g. \citealt{ANN_redshifts}).

Due to the significant inherent uncertainties in such redshift estimates, for analysis objects may be grouped into bins of similar redshift. In order to achieve the precisions desired for current and future generations of cosmological experiments, we need to be able to determine the redshift distribution of each of these bins with greater accuracy than has been possible by simply using standard fitting to spectroscopic data. The impact of the redshift distributions on cosmology, and the importance of knowing them to a high degree of accuracy, has been the subject of a number of studies such as \cite{Huterer_needs}, \cite{Newman_CC}. These suggest that to achieve the desired precision and accuracy in upcoming experiments such as LSST, we require that the mean and width of redshift bins to be known to \m{O(10^{-3}(1+z))}. \cite{Newman_CC} proposes that this may be achieved by calibrating the the photometric redshifts using information from cross-correlations with spectroscopic data.

Since the proposal of these ideas, there have been some studies looking into the potential for using cross-correlations for estimating photometric redshift distributions (e.g. \cite{Menard_Calibration}, \cite{Schmidt_Recovery}, \cite{Schulz_Calibration}, \cite{Matthews_Reconstruction}, \cite{McQuinn}), as well as potential problems such as contamination as in \cite{Benjamin_Contamination}. These tend to focus on recovering the redshift distribution from simulations by comparing the correlation between some photometric data set and a spectroscopic sample at known redshift, and assuming some fixed cosmology. In the case of a practical analysis however, we will not know the cosmological parameters (the determination of which is, after all, the objective of such calculations), and the calculation of theoretical correlation functions is cosmology dependent. It is well known that the cosmological parameters and redshift distribution are degenerate, and hence we cannot estimate how well the redshift distribution can be constrained without also varying the cosmology itself, as uncertainties from the cosmology may become a significant factor. This may be particularly important where the region of overlap between the photometric sample and the spectroscopic sample is relatively small. Hence, in order to avoid biases or overly optimistic estimates of our constraining power, we must determine the cosmology and the redshift distribution together, rather than treating them as independent problems. Previous work such as  \cite{Newman_CC} and \cite{McQuinn} also use estimators which may be prone to finding local maxima, and do not explore the space as fully as an MCMC approach using a full likelihood. 

A significant amount of attention in recent years has been placed on the power of cross-correlations as a statistical tool for cosmology (such as \cite{Rhodes_CC}, \cite{CrossCorr}). Using cross-correlations to calculate both redshift distributions and cosmological parameters implies that we may include these effects into one framework with relative ease. We demonstrate such a technique for calibrating the photometric redshift distribution from an initial estimate using a joint likelihood analysis with cosmology using the angular power spectrum \m{C(l)}. We take the errors in photometric redshift modelling into account by allowing the mean and width of the photo-z bins to vary as free parameters, just as we do with cosmological parameters. The redshift binning is then marginalised over in order to obtain the probability contours for cosmological parameters. This allows us to study the extent to which we can constrain the photometric redshift distributions and simultaneously explore the impact of this information on cosmological inferences, in a manner which automatically treats the errors in the distributions in a Bayesian way. 

In this paper we investigate extent to which photometric redshift bins can be constrained by cross correlations, and the impact on cosmological parameter inference in the case of large scale structure. We present a simplified experiment where we vary the width and mean of gaussian redshift bins, although we explain how the framework may be applied to higher moments also. For computational simplicity, only three photometric and three spectroscopic bins will be used (although this can be extended to fuller surveys at the expense of computation time); this simple model should suffice to demonstrate the power of the technique, as well as the degeneracies between the parameterisations of the redshifts and the cosmological parameters. The impact on future optical surveys will be greater though, as the same technique can be used to constrain photometric samples in weak lensing analyses, which may be used in conjunction with galaxy number counts to infer cosmology. 

\section{The C(l) Calculation} \label{Cls}

The angular power spectrum is split up into correlations between different bins and cosmological probes; the full object we wish to look at is \m{C^{ij}_{\alpha \beta}(l)} where \m{i,j} vary over labels of bins and \m{\alpha, \beta} vary over cosmological probes such as galaxy number counts or shear measurements. (Indices may be suppressed when they are not relevant.) \m{C^{ij}_{\alpha \beta}(l)} is symmetric in \m{i ,j} and \m{\alpha, \beta}. 

\subsection{The C(l) formalism}

Following the approach of \cite{Peebles} -- and later \cite{Blake_1}, \cite{Thomas} -- for a particular probe of our cosmology observed projected on the sky in the direction of a unit vector \m{\b{n}}, \m{X(\b{n}) = \bar{X} + \Delta X(\b{n})}, we may decompose the variation in this parameter \m{\Delta X} into spherical harmonics as
\begin{equation}
\Delta X(\b{n}) = \sum\limits_{l>0} \sum\limits^l_{m=-l} a_{lm} Y_{lm}(\b{n})
\end{equation}
We may calculate the coefficients \m{a_{lm}} by using the orthogonality of spherical harmonics (\m{\int Y_{lm}Y^*_{l^\prime m^\prime} d\Omega = \delta_{ll^\prime}\delta_{mm^\prime} }) :
\begin{equation}
a_{lm} = \int \Delta X(\b{n}) Y^*_{lm}(\b{n}) d\Omega 
\end{equation}
The \m{C(l)}s are defined from these coefficients by the relation:
\begin{equation}
C(l) = \langle a_{lm} a^*_{lm} \rangle
\end{equation}
In our case we are interested in the galaxy distrtibution as a tracer of matter; this is calculated from the data by analysing number counts across the sky. It is important to note that this does not require knowledge of \m{n(z)}: we do not use redshift information in calculating the angular power spectrum from the data. 

For the theoretical modelling however, we do require knowledge of \m{n(z)}, as we must calculate the full power spectrum \m{P(k,z)} which is then projected onto the sky. This projection, as we shall later see, is strongly dependent on \m{z}.  To calculate the \m{C(l)}s we use the following equation (\citealt{Thomas}):
\begin{equation} \label{limber}
C^{ij}_{\alpha \beta}(l) = \frac{2}{\pi} \int W^i_\alpha(l,k) W^j_\beta(l,k) k^2 P(k) dk 
\end{equation}
although we will perform most calculations without this approximation, it is useful to understand the impact of the redshift distributions. The redshift distributions \m{n^i(z)} enter the \m{C(l)}s through the window functions. 


\subsection{Window functions} \label{windows}
Window functions allow us to project the distribution of galaxies onto the sphere and decompose into spherical harmonics. Here we will discuss only the window function for galaxy clustering, since we have not used other probes in this particular work.

\subsubsection{The galaxy clustering window function}

\cite{Huterer_Cl} derive a calculation for galaxy clustering information of the following form.
\begin{equation} \label{window_n}
W^i_{\text{g}}(l,k,z) = \int b_g(k,z) n^i(z) j_l(k \chi) D(z) 
\end{equation}
Here \m{n^i(z)} is the redshift distribution in bin \m{i}, \m{b_g(k,z)} is the galaxy bias, \m{D(z)} is the growth function, and \m{j_l(k \chi)} is the order \m{l} spherical bessel function. 
Note that the comoving distance to an object is a function of redshift \m{\chi(z)}. 

\subsubsection{Including Redshift Space Distortions}
Redshift Space Distortions (RSD) are alterations to the redshift of a galaxy due to its peculiar velocity. This leads to a distortion of the galaxy distribution if we attempt to reconstruct the three dimensional information, with galaxies with peculiar velocity toward us appearing closer (at lower redshift) and galaxies with peculiar velocity away from us along the line of sight appearing further (at higher redshift). Since these peculiar motions are due to interactions with local gravitational potentials they contain cosmological information. RSD on linear scales can be included by an additional term in the window function, following \cite{CrossCorr}.

\begin{multline}
	W^i_{\text{RSD}}(l,k,z) = \beta \int \phi(\chi) [ \frac{2l^2 + 2l -1}{(2l+3)(2l-1)}j_l(k\chi) \\ - \frac{l(l-1)}{(2l-1)(2l+1)}j_{l-2}(k\chi) - \frac{(l+1)(l+2)}{(2l+1)(2l+3)}j_{l+2}(k\chi) ] d\chi
\end{multline}
The complete window function to be used in our \m{C(l)} calculation is then given by the sum of these two terms.
\begin{equation}
	W^i_{\text{LSS}} = W^i_g(l,k) + W^i_{\text{RSD}}(l,k)
\end{equation}

In this paper we will not consider the effects of galaxy bias, although some papers have noted the potential importance of evolving galaxy bias in determining redshifts from correlation data (\cite{Schmidt_Recovery}, \cite{Schulz_Calibration}). In the absence of a compelling bias model however, bias may be best handled as nuisance parameter (or parameters) which is also marginalised over. This is demonstrated in \cite{Clerkin_bias}, but for simplicity we choose a constant bias \m{b_g = 1}. 

We can now see how the redshift distribution enters into the \m{C(l)} formalism. If we have an accurate redshift distribution, for instance from a spectroscopic survey, then this is all we need to begin calculating our theoretical correlations. Unfortunately, photometric estimates are far from perfect, and photometric redshift errors if left unignored may produce unforeseen effects in our computed \m{C(l)}s. We will now seek to understand what some of these effects may be. 

\subsection{The significance of n(z)} \label{overlaps}

Intuitively one might expect that number counts on the same patch of sky will be highly correlated when close in redshift and less correlated when widely separated. If we have a spectroscopic sample and a photometric sample that overlap in redshift, then they will contain objects in the same larger clustering structures, which we will be able to see as boosts in their correlations. We do not expect to see clustering over very large distances, so we expect that samples widely separated in redshift will show very weak cross-correlations.

We can put this intuitive understanding on a more mathematical foundation. From the definition of the window functions and the C(l) calculation, we can see how we expect \m{n(z)} to affect our calculated \m{C^{ij}(l)}. If, for the sake of simplicity, we assume a \m{k}-independent bias \m{b_g(z)}, then the \m{k} dependence of \m{W(l,k)} comes entirely from the spherical Bessel function \m{j_l(k \chi)}. The window functions oscillate as a function of \m{k}, made of contributions with different frequencies set by the spherical bessel functions in the integral. Hence the redshift range of the integral sets the range of frequencies present in the window function. If the distribution for a particular bin \m{n^i(z)} is close to zero outside a particular range (for instance, if we model \m{n(z)} as a top hat or Gaussian function) then the integral over \m{z} has a fairly small range which contributes significantly. If two bins \m{n^i(z)} and \m{n^j(z)} are separted in \m{z} by significantly more than their variance, then our two window functions \m{W^i(l,k)} and \m{W^j(l,k)} will have only very small contributions with the same frequency. The product of two oscillating functions with different frequencies will tend to average to zero when integrated over, so we would expect that the integral over these two window functions to be small. If, however, the redshift ranges overlap in regions of significant number density, then there will be significant contributions to both window functions with the same frequency and forms. These, when integrated over, will not average to zero and give a large contribution to \m{C^{ij}(l)}. Hence we expect the \m{C^{ij}(l)}s to be dependent on the amount of overlap between distribution functions in different bins, with significant overlaps in areas with high number density giving the strongest signals. 

In addition to the overlap between bins, the spherical bessel function in equation \ref{window_n} also tells us more about the redshift dependence of \m{C^{ij}(l)}: we expect stronger signals from distributions at lower redshift where the amplitude of \m{j_l(k\chi(z))} is higher. So whilst the overlap between bins will determine the relative power in cross-correlations compared to auto-correlations, moving all the bins together up or down in redshift can shift the amplitudes of all the signals together. The redshift distribution is of course not the only thing which will affect our signal, and cosmological effects enter into our equations through the growth function (\m{D(z)} in equation \ref{window_n}) and the power spectrum (\m{P(k)} in equation \ref{limber}). This is the source of a very important degeneracy between our redshift distributions and comsological parameters, particularly those such as \m{A_s} or \m{\sigma_8} which strongly control the amplitude of \m{P(k)}. 

In order to fix the redshift distribution, we need bins which overlap our photometric redshifts but are strongly anchored so that any changes in photometric bins, even moving coherently, will be captured by the \m{C(l)}s. For this we require spectroscopic data, which is well known enough to have rigidly fixed \m{n(z)}, which overlaps our photometric bins. Note that our only criterion here is that our spectroscopic and photometric data overlap, and not - unlike with template and machine learning techniques - that the spectroscopic sample be an unbiased representation of the photometric sample. 

\section{Modelling the Redshift Distributions} \label{bins}

\subsection{Photmetric redshifts} \label{photoz}

Due to the uncertainty in photometric redshifts, we cannot obtain an accurate redshift for each object that we have in our sample. We instead model a bin as a broader function which captures the distribution of redshifts which would be binned together. 

In this work we will model photometric redshift distributions within a bin as a Gaussian distribution defined by their mean and variance. 
\begin{equation}
n(z,\mu,\sigma) = G(z, \mu, \sigma) = \left(\frac{1}{2\pi \sigma^2}\right)^{\frac{1}{2}} \exp \left[- \frac{ (z-\mu)^2}{2\sigma^2} \right] 
\end{equation}

In order to model the uncertainties in \m{n(z)}, we need to be able to control the shape of the function in a quantitive way, ideally with as few parameters as possible. Each time we add a parameter, we are adding \m{N_{bins}} new dimensions to our parameter space to be explored by the MCMC and hence our computation becomes exponentially more expensive. The most important parameters are the mean and width of the distribution; other adjustments to the shape can be abandoned without too much impact but nevertheless the method is general and in wider, strongly non-Gaussian redshift bins higher moments may be taken into account if necessary. When using the mean and variance of a gaussian distribution we may adjust $\mu$ and $\sigma$ directly using the analytic formula for a Gaussian. To vary a general distribution, or to change the shape in other ways, you may refer to the appendix.

Although this is the template used for all the bins in this study, we may also apply non-Gaussian distortions to these distributions to model more complex effects. It is  also important to note that this method is by no means limited to Gaussian functions, and these may be easily replaced by an arbitrary function (with some parametrisation) \m{F(z, \textbf{p})}, where \m{\textbf{p}} is the parameter vector to be marginalised over. In the simple example above \m{\textbf{p} = (\mu,\sigma)}, although we may extend this to include skew and kurtosis, and have \m{\textbf{p} = (\mu,\sigma,s,k)} or some other vector of parameters. In a typical survey such as DES, photometric redshift bins have a standard deviation of approximately 0.1 to 0.2 (\citealt{DESForecast}); for our purposes we will take \m{\sigma = 0.1}. 

\subsection{Spectroscopic redshifts} \label{specz}

In order to include spectroscopic redshifts into the same formalism, we model spectroscopic redshifts in bins with much narrower distributions. This will take the form of a much narrower Gaussian. (A narrow top-hat function may also be used, but smooth continuous functions are often computationally more stable.) We assume that spectroscopic information is known well enough that we do not vary these bins in the same way as the photometry, and so there is no parameter vector \m{\textbf{p}} to marginalise over. Spectroscopic bins will be modelled with a width of \m{\sigma = 0.025}, which requires spectroscopic redshifts to be estimated to within a few percent. Their thickness may be determined by the nature and quality of the spectroscopic sample, or as a compromise with computational efficiency. A small number of wider bins is less computationally expensive than many narrow bins; the width of spectroscopic bins makes little to no impact on the length of computation (integrations are performed between fixed redshifts), however each additional bin adds two new parameters, which means that we have more integrations to perform (scaling as \m{N_\text{bins}^2}), larger covariance matrices, and a much larger parameter space which scales exponentially in volume with the number of parameters. Narrow bins allow us to look at very localised correlations at the cost of this additional computation. 

\section{The Likelihood Function and MCMC Methods} \label{likelihood}

\subsection{The likelihood function for C(l)s} 

The likelihood is calculated from \m{C(l)} for each model (i.e. each parameter set) compared to the \m{C(l)} calculated from the fiducial model. Let the fiducial model be known as model A, and the model we wish to investigate model B; we calculate a log-likelihood of seeing some fluctuations \m{a_{lm}} in the model B compared to the model A, and take then take an expectation value assuming the fiducial model A in the absence of any data, as described in \cite{ProbRef}. Since \m{a_{lm}} are stochastically generated, any given cosmology may generate a wide variety of \m{a_{lm}}, and each set of \m{a_{lm}} can give therefore give a different likelihood when compared against a model. Hence, with no reason to generate one particular set over another, one calculates the expectation value of these possible likelihoods, on the assumption that our \m{a_{lm}} were generated by the cosmology represented by A. This quantity is dependent only on the \m{C(l)}s calculated in each model, and the properties of the survey such as sky coverage and noise which remain constant throughout.

For measured \m{a_{lm}}, using the fact that the expectation value is zero, we have for a given cosmology X the relation
\begin{equation} \label{alm_variance}
\text{Var}(a_{lm}) = \langle |a_{lm}|^2 \rangle_X = C_X(l) + N(l) 
\end{equation}
where noise is assumed to be isotropic and uncorrelated (shot noise) and taken into account by the noise function \m{N(l)}. Assuming Gaussian distributions, we then have 
\begin{equation} \label{alm_gauss}
P(a_{lm} | X) =  \left( \frac{1}{2\pi (C_X(l) + N(l)) } \right) ^{\frac{1}{2}} \exp \left[- \frac{|a_{lm}|^2}{2(C_X(l)+N(l))} \right] 
\end{equation}
We wish to calculate the (expected) likelihood function 
\begin{equation} \label{log_PaPb}
\left<L\right> = \left< \log \left[ \frac{P(a_{lm} | B)}{P(a_{lm} | A)} \right] \right>_A 
\end{equation}
Given equation \ref{alm_gauss} we can write 
\begin{multline} \label{PA_PB}
\frac{P(a_{lm}|B)}{P(a_{lm}|A)} = \left[ \frac{C_A(l) + N(l)}{C_B(l) + N(l)} \right] ^{\frac{1}{2}} \\ \exp \left[ \frac{|a_{lm}|^2}{2(C_A(l) + N(l))} - \frac{|a_{lm}|^2}{2(C_B(l) + N(l))} \right]
\end{multline}
Taking logs we obtain
\begin{multline} \label{logPA_PB}
\log\left(\frac{P_A}{P_B}\right) = \frac{1}{2}\log\left( \frac{C_A(l) + N(l)}{C_B(l) + N(l)}\right) \\ + \frac{|a_{lm}|^2}{2(C_A(l)+N(l))} -  \frac{|a_{lm}|^2}{2(C_B(l)+N(l))} 
\end{multline}
We then take the expectation value assuming A using equation \ref{alm_variance}
\begin{equation} \label{PA_PB_A}
L = \frac{1}{2} \left[1 - \frac{C_A(l) + N(l)}{C_B(l) + N(l)} +  \log \left(\frac{C_A(l) + N(l)}{C_B(l) + N(l)}\right) \right]
\end{equation}
We must then take into account all of the \m{a_{lm}}, bearing in mind this expression is not dependent on \m{m}, so we have \m{2l + 1} identical terms for each \m{l}, and taking into account the fraction of the sky \m{f_{s}} observed.
\begin{equation} \label{PA_PB_A}
L =  \frac{f_s}{2} \sum\limits^{l_{max}}_{l=2} (2l+1) \left[1 - \frac{C_A(l) + N(l)}{C_B(l) + N(l)} +  \log \left(\frac{C_A(l) + N(l)}{C_B(l) + N(l)}\right) \right]
\end{equation}
This is the likelihood we will use for CMB temperature information (\m{C^{TT}(l)}) or when we calculate the autocorrelation of a bin. When we have multiple bins or cosmological probes where cross correlations must be taken into account, then we will have more than one \m{C(l)} function for each cosmology. We then a multivariate Gaussian distribution instead of simply the product of independent Gaussians. The covariance matrices are
\begin{equation} \label{matrices}
\left[ M_{X,l} \right]_{i,j} = C_{(X)}^{ij}(l) + \delta^{ij} N^i(l)
\end{equation}
where X may be either A or B (with the relavent \m{C(l)}s calculated in the left hand side) and where \m{N(l)} is the noise associated with the experiment. Noise is only added on the diagonal as shot noise between bins should not be correlated and hence not contribute to the covariance. This gives a probability distribution
\begin{equation}
P( \textbf{a}_{lm} | X ) = \frac{1}{\left( (2\pi)^k |\textbf{M}_{X,l}| \right)^{\frac{1}{2}}} \exp \left[-\frac{1}{2} \textbf{a}_{lm}^T \textbf{M}_{X,l}^{-1} \textbf{a}_{lm} \right]
\end{equation}
for a \m{k \times k} covariance matrix (i.e. cross correlating $k$ bins). Repeating the above analysis, and using the following relations (where \m{l,m} subscripts have been suppressed for clarity, and we use summation convention over \m{i,j})
\begin{multline}
\left< \textbf{a}^T \textbf{M}^{-1}_{X} \textbf{a} \right>_A = \left< a_i M^{-1}_{X,ij} a_j \right>_A = M_{A,ij} M^{-1}_{X,ij}
\\ = M_{A,ji}M^{-1}_{X,ij} = \left[ \textbf{M}_A \textbf{M}_{X}^{-1} \right]_{jj} = \text{Tr}\left[ \textbf{M}_A \textbf{M}_{X}^{-1} \right]
\end{multline}
we arrive at the analagous log-likelihood to equation \ref{PA_PB_A} for multiple \m{C(l)}s
\begin{equation} \label{like}
L = \frac{f_{s}}{2} \sum\limits_{l} (2l + 1) \left[ \text{Tr} \left( I-M_{A,l}M_{B,l}^{-1} \right) + \ln \left(\det \left( M_{A,l}M_{B,l}^{-1} \right) \right) \right]
\end{equation}
It is easy to see that this is zero for A=B.  

\subsection{Noise parameters and survey assumptions} \label{survey_noise}

\subsubsection{Galaxy number counts}
We limit our model to shot noise, which is described by the noise function
\begin{equation} \label{noise}
N^i(l) = (\sigma^i(l))^2 = \frac{1}{\bar{n}^{i}} = \frac{ 4 \pi f_{s} }{ f^i_g N_g } 
\end{equation}
Where \m{f_{s}} is the fraction of the sky observed by the survey, \m{f^i_g} is the fraction of the total number of galaxies observed which lie within that redshift bin \m{n^i(z)}, and \m{N_g} is the total number of galaxies observed over the entire survey. For a DES like survery we assume that \m{N_g = 3 \times 10^8}, \m{f_{s} = 0.12} (from \m{A = 5000 \text{deg}^2}), and a redshift range \m{0 < z \le 2} (\citealt{DESForecast}).

\subsubsection{CMB TT information}
Here we have a slightly more complex function which must take into account more survey information. We base our parameters on a Planck-like survey, based on the parameters described in \cite{Fil_Prob}. 
Our noise function is 
\begin{equation}
N^2_l = \sum_{chan} \frac{1}{(\sigma_c \theta_b)^2} \exp\left(-\frac{l(l+1)\theta_b}{8\ln 2}\right)
\end{equation}
\begin{equation}
\sigma_c = \frac{T_{\text{NE}} \theta_{\text{sky}}}{\sqrt{n_{\text{det}} t}\theta_b}
\end{equation}
where \m{\theta_b} is the beam width, \m{T_{\text{NE}}} is the noise effective temperature, \m{n_{\text{det}}} is the number of detectors, and \m{t} is the integration time assumed to be one year. We assume information is collected in four bands with parameters detailed in Table \ref{CMB_params}. 
For CMB information we assume \m{f_{\text{sky}} = 0.65}.

\begin{table}
\centering
\caption{Parameters for a Planck like CMB survey.}
\label{CMB_params}
\begin{tabular}{|l|l|l|l|l|}
\hline
Band Frequency       & 70   & 100 & 143 & 217 \\ \hline
Beam Width $\theta_b$  / arcsec   & 14.0 & 9.5 & 7.1 & 5.0 \\ \hline
Noise Effective Temperature / $\mu K \sqrt{s}$           & 212  & 56  & 56  & 84  \\ \hline
Detector Number $n_{\text{det}}$ & 12   & 8   & 12  & 12  \\ \hline
\end{tabular}
\end{table}

\subsection{Computational details: UCLCl and PLINY codes} \label{codes}

\m{C(l)} calculations are performed using the UCLCl code developed at UCL, and the CLASS Boltzmann code (\citealt{CLASS}) for the generation of the primordial power spectrum and transfer function. Within UCLCl most functions, including \m{n(z)}, are represented using splines. The spline representation is advantageous for this work because it allows us to easily manipulate and deform \m{n(z)} in non-linear ways without having to define an analytic function with some parametrisation. (This means we could take an arbitrary form from, for instance, data and still manipulate it in the way described in this paper.) We can vary the mean and variance for an arbitary distribution in a precise way. For the higher moments such as skew and kurtosis, we must vary these more heuristically for a general distribution, and these transformations may affect other moments. These may all be varied by applying transformations to the \m{z} variable of the \m{n(z)} spline, as discussed in the appendix.

The MCMC analysis is performed using PLINY, a nested sampler designed for parallel computation. It calculates a chain of points in the parameter space, with likelihoods and prior weights, and also outputs an evidence calculation. In order to calculate the posterior weight for each point in the chain we need to use Bayes' Theorem:
\begin{equation} \label{Bayes}
\text{Posterior} = \frac{ \text{Likelihood} \times \text{Prior} }{\text{Evidence}}
\end{equation}
For all parameters in this analysis we assume flat priors with hard edges well away from the peak of the distribution. 
The evidence is not strictly necessary in this analysis as it is just a constant factor. The evidence is only required if we wish to perform a model comparison for models with different parameterisations. 

\subsection{The fiducial model} \label{fiducial}

In this work we use a fiducial \m{\Lambda}CDM model. For the sake of computational efficiency, we take work only with flat cosmologies (\m{\Omega_k = 0}). We also restrict ourselves to varying seven cosmological parameters - \m{\{A_s, \Omega_\Lambda, \Omega_b, h, n_s, \tau_{r}, w_0 \}}. Our fiducial cosmology will be 
\begin{multline}
A_s = 25\times10^{-10}, \\ \Omega_\Lambda = 0.7, \\ \Omega_b = 0.06,\\ h = 0.7,\\ n_s = 0.95, \\ \tau_{r} = 0.09, \\ w_0 = -0.9 \\
\end{multline}
To speed computation we limit ourselves to flat cosmologies, and hence we will use \m{\Omega_{\text{cdm}} = 1 - \Omega_\Lambda - \Omega_b}, which gives a fiducial \m{\Omega_{\text{cdm}} = 0.24}. 
We use three photometric bins with mean
\begin{equation*}
(\mu_1, \mu_2, \mu_3) = (0.8, 1.0, 1.2)
\end{equation*}
and standard deviation 
\begin{equation*}
\sigma_1 = \sigma_2 = \sigma_3 = 0.1
\end{equation*}
For analyses with spectroscopy, we use three spectroscopic bins with mean
\begin{equation*}
(\mu_1, \mu_2, \mu_3) = (0.7, 1.0, 1.3)
\end{equation*}
and standard deviation 
\begin{equation*}
\sigma_1 = \sigma_2 = \sigma_3 = 0.025
\end{equation*}
Spectroscopic bins are assumed to be well known enough not to need variation in the MCMC analysis, so this leaves us with an 13-dimensional parameter space (7 cosmological and 6 photometric binning parameters). 

For our noise function, in this particular analysis, we have chosen DES like parameters (described in section \ref{survey_noise}) with \m{f_{s} = 0.12}, \m{f^i_g = 0.2} for all photometric bins, and \m{N_g = 3 \times 10^8} (giving \m{6\times10^7} galaxies in each photometric bin). For spectroscopic bins, we assume that we have \m{5\times10^4}  galaxies in each bin, roughly in keeping with the density in surveys such as BOSS or eBOSS (\citealt{BOSS}, www.sdss.org). 

For the sake of reasonably rapid calculations we use a limited number of redshift bins in this demonstration, although most full surveys will use 5-10 photometric redshift bins. This will diminish our power to constrain the cosmological parameters somewhat due to a lack of information and coverage over much of the redshift range, but will be enough to demonstrate the power of the technique applied to the calibration of photometric redshifts. 

\section{Results} \label{results}

In this section we will present the results of our MCMC analysis. We will first demonstrate the bias in cosmological parameters that is caused by having poorly estimated photo-z bins. We will then show what can be achieved using autocorrelations of photometric bins, where bins are allowed to vary freely; this will demonstrate where the degeneracies between photometric redshifts and cosmological parameters lie. Finally we will show results using cross correlations between both photometric and spectroscopic bins, which gives dramatically improved precision on the photometric bins, and we demonstrate the effect of this on the marginalised distributions for the cosmological parameters. 

\subsection{Cosmological parameter bias from n(z)} \label{bias}

From equations \ref{limber} \& \ref{window_n}, our theoretical prediction of \m{C^{ij}(l)} is dependent on the redshift distributions \m{n^i(z)} and \m{n^j(z)}. If we estimate properties of our redshift bins (in this case \m{\mu}, \m{\sigma}) by fitting objects with known spectroscopic redshifts, then we will derive redshifts with some scatter around their `true' value. These redshift errors have a knock-on effect on our inference of cosmological parameters. For example, if our estimated redshifts are too low, then theoretical power that we calculate will be too high; in order to match the observations, \m{A_s} may be lowered to match the power, and other parameters adjusted to get the best fit to shape. In this section we will demonstrate such biases, and later we shall see how marginalising over redshift distributions can avoid them. Methods in estimating photometric redshifts often have an error in \m{z} of \m{O(0.1)}, which is large compared to what we would require to obtain precise results from a photometric survey. If take the parametrisation obtained from this fitting on face value then we will reconstruct a slightly distorted \m{n(z)}. This means that when we fit our cosmological model, our cosmological parameters will be inevitably be changed in order to counter the effect of the distortions in \m{n(z)}. We may analyse this case in our simple model by using the fiducial \m{n(z)} for the "observed" \m{C(l)}s as described in section \ref{fiducial}, but calculating our model \m{C(l)}s using bins fixed to have different parameters. 

\begin{figure} 
	\includegraphics[width=8cm,keepaspectratio=true]{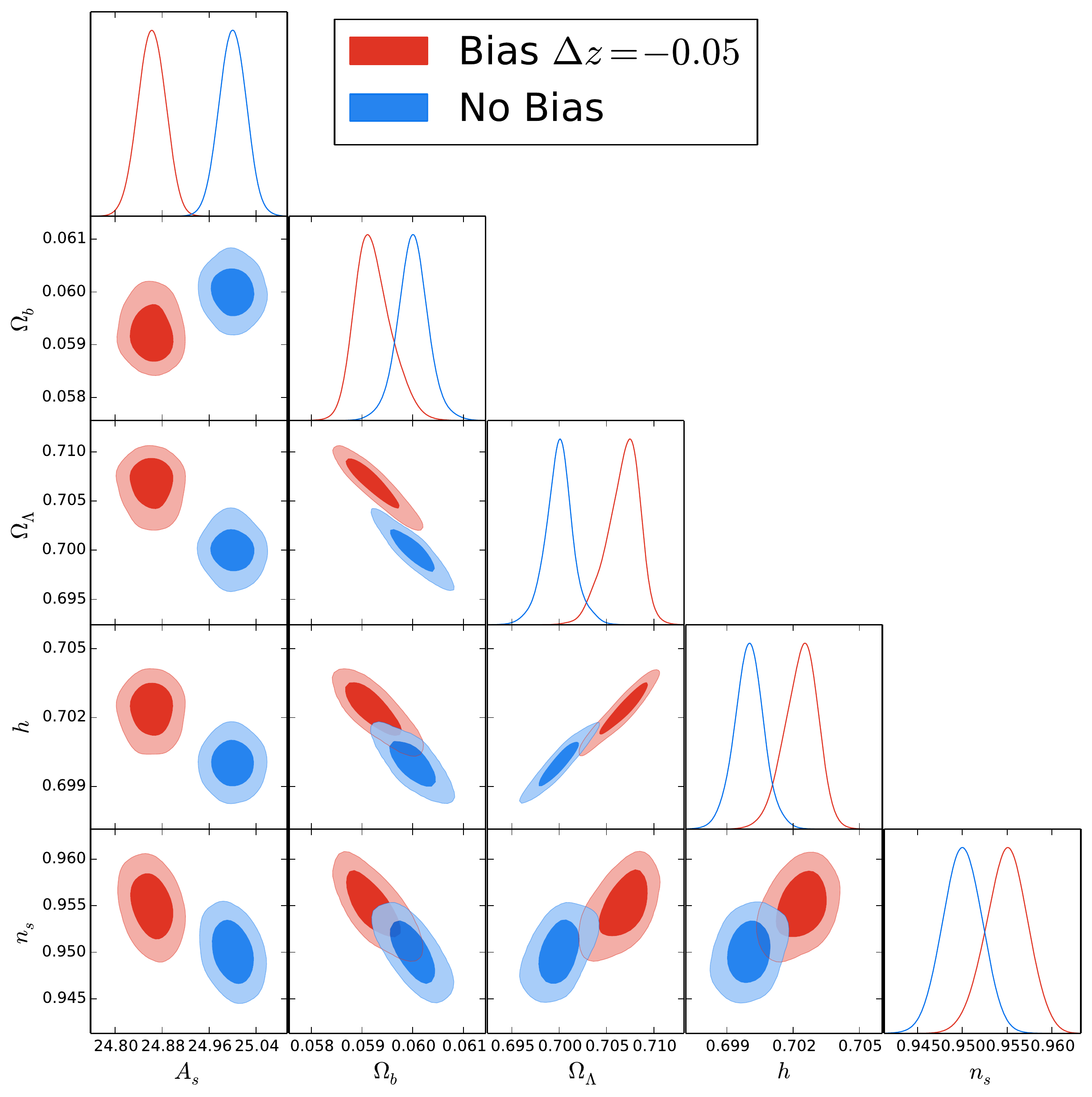}
	\caption{Probability contours obtained from the true redshift distribution (shown in blue) and from a biased redshift distribution (shown in red). A five parameter cosmology is derived from two photometric bins, where the fiducial cosmology and the blue contours use $\mu_1 = 0.8$ and $\mu_2 = 0.9$, whereas the red contours are derived on the incorrect assumption that $\mu_1 = 0.75$ and $\mu_2 = 0.85$ i.e. photometric redshifts are systematically underestimated. The blue contours are, by construction, centred on the fiducial parameters, whereas the red contours end up far from the fiducial parameters in order to compensate for effects in the $C(l)$ signal introduced by photometric systematics.}
		\label{bias_triangle}
\end{figure}

The cosmology used is the same as stated in section \ref{fiducial} but we shall only use two redshift bins at \m{\mu_1 = 0.8} and \m{\mu_2 = 0.9} to generate the fiducial \m{C(l)}s. When we attempt to recover the cosmological parameters with an MCMC analysis, we use a fixed redshift distribution, biased with \m{\mu_1 = 0.75} and \m{\mu_2 = 0.85}. The results are shown in Figure \ref{bias_triangle}.

The bias is strongest in cases such as this where there is a systematic error causing the mean or standard deviation of bins to be consistently over or under estimated. In order to avoid this, we must reduce our reliance on fixed redshift distributions with large errors. In lieu of a method for sufficiently accurate redshifts from photometry, we must rely on marginalising in a bayesian framework, the results of which are described in the following sections. 

\subsection{Autocorrelations with photometric redshift bins}

\begin{figure} 
	\includegraphics[width=8cm,keepaspectratio=true]{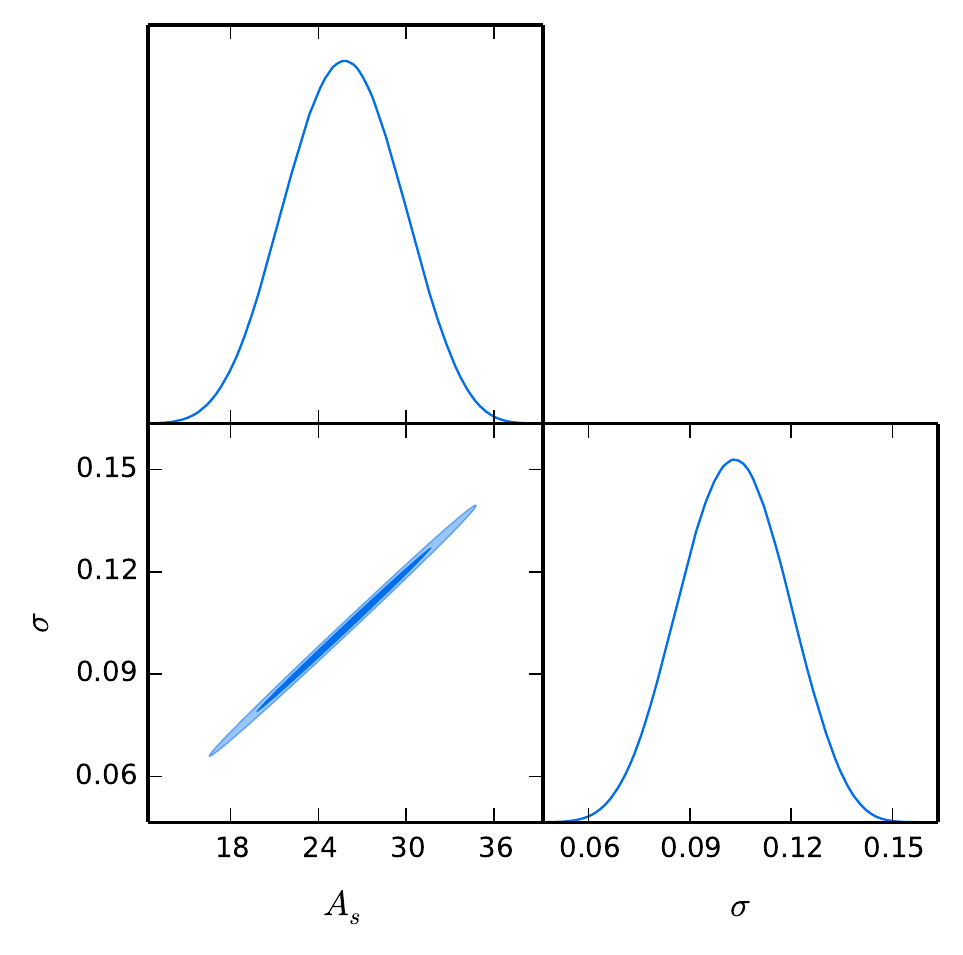}
	\caption{Probability contour obtained when varying only $A_s$ and $\sigma$ (the width of one photometric redshift bin), with $\mu = 1.0$ and all other parameters at their fiducial values, demonstrating the high degeneracy between photometric bin width and the $A_s$ parameter.}   \label{Asigma}
\end{figure}

The simplest analysis that we can do is to use only our photometric redshift bins, and to only take into account autocorrelations. We will see that this means ignoring a great deal of information, and our bounds on cosmological and binning parameters are wide. Although in this case we are not taking into account the full information available to us, it is worth looking into since it is much less computationally expensive, and previous studies have been focussed on autocorrelations. We promote the mean and standard deviation of our redshift bins to fully independent parameters for our MCMC analysis, allowing them to vary freely so that they can be marginalised over. Here we use the three photometric bins described in section \ref{fiducial}. This information is combined with CMB TT information in order to help constrain \m{A_s}, which is a problematic parameter in this analysis since it is extremely strongly degenerate with the standard deviation of a bin (see Figure \ref{Asigma}). 

\begin{figure*} 	\includegraphics[width=17.5cm,keepaspectratio=true]{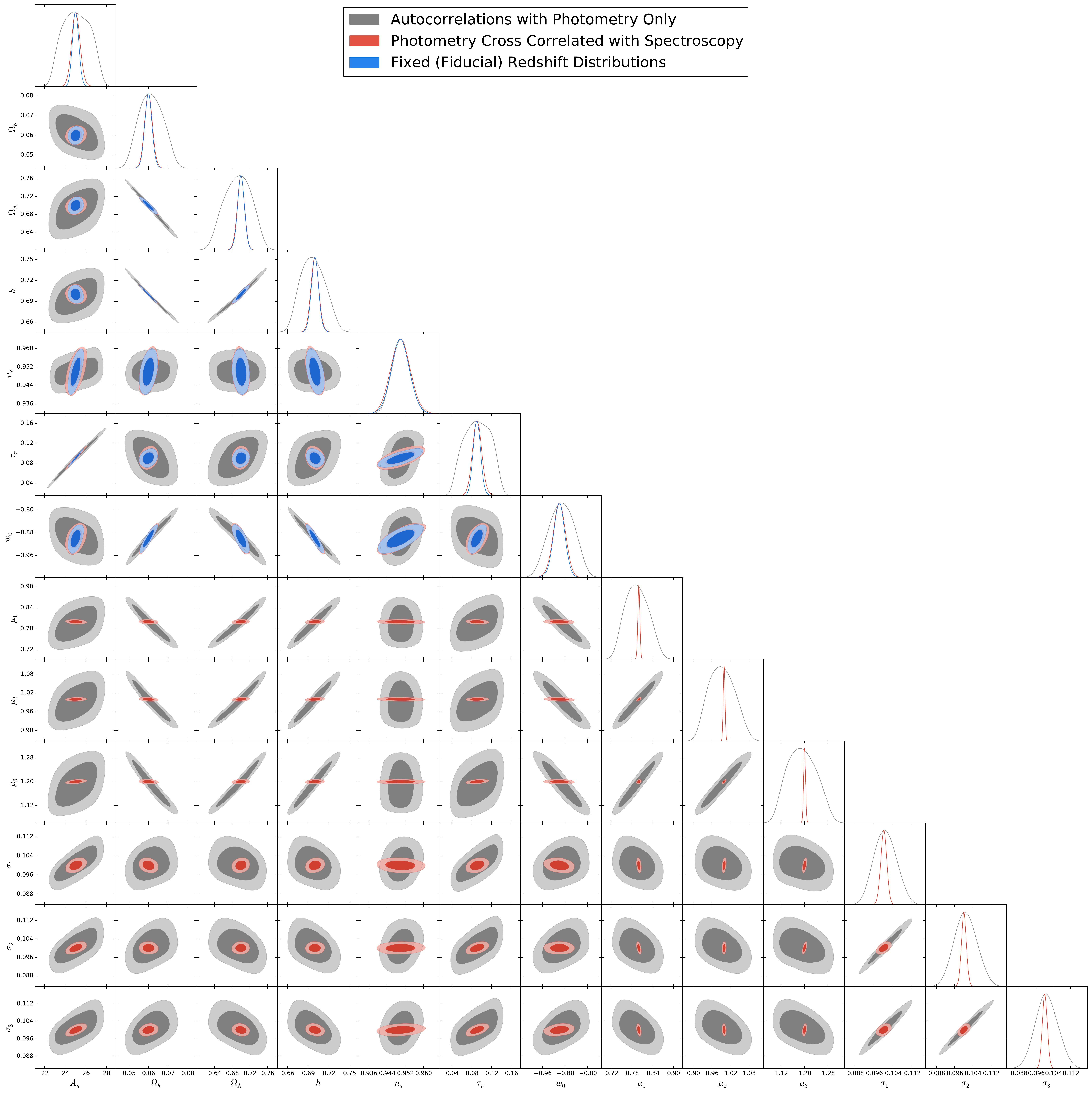}
	\caption{Probability contours obtained for cosmological and photometric binning parameters. Results from using autocorrelations of photometric bins only are shown in grey, and results from using photometric bins cross correlated with each other and spectroscopic bins are shown in red. For comparison, results using redshift bins fixed at the fiducial values are shown in blue. Results show clear improvements on all the binning parameters, as well as most cosmological parameters (with the exception of $n_s$), and contours using cross correlations between photometric and spectroscopic samples yield results very close to those with no redshift error.} 	\label{Triangle}
\end{figure*}
\begin{table*}
\centering
\caption{68\% confidence ranges for inferred cosmological parameters using only autocorrelations, and using cross correlations with spectroscopy.}
\label{cosmo_table}
\begin{tabular}{|l|l|l|l|l|l|l|l|}
\cline{2-8}
             & $A_s\times10^{10}$ & $\Omega_b$    & $\Omega_w$  & $w_0$         & $h$         & $\tau_r$      & $n_s$         \\ \hline
Fiducial     & 25                 & 0.06          & 0.7         & -0.9          & 0.7         & 0.09         & 0.95          \\ \hline
Photometric  & $25.1 \pm 1.5$  & $0.060 \pm 0.007$ & $0.693 \pm 0.033$ & $-0.89 \pm 0.05$ & $0.697 \pm 0.019$ & $0.091 \pm 0.029$  & $0.950 \pm 0.004$ \\ \hline
Photo $\times$ Spec & $25.0 \pm 0.4$     & $0.060 \pm 0.002$ & $0.700 \pm 0.008$ & $-0.90 \pm 0.02$ & $ 0.700 \pm 0.006 $ & $0.091 \pm 0.009$ & $0.950 \pm 0.004 $\\ \hline
Fixed Redshift & $25.0 \pm 0.3$     & $0.060 \pm 0.002$ & $0.700 \pm 0.008$ & $-0.90 \pm 0.02$ & $ 0.700 \pm 0.005 $ & $0.090 \pm 0.008$ & $0.950 \pm 0.004 $\\ \hline
\end{tabular}
\end{table*}

\begin{table*}
\caption{68\% confidence ranges for inferred photometric redshift bin parameters, using only autocorrelations and using cross correlations with spectroscopy.}
\label{bin_table}
\begin{tabular}{l|l|l|l|l|l|l|}
\cline{2-7}
                               & $\mu_1$         & $\mu_2$       & $\mu_3$      & $\sigma_1$   & $\sigma_2$   & $\sigma_3$  \\ \hline
Fiducial & 0.8         & 1.0         & 1.2         & 0.1         & 0.1         & 0.1         \\ \hline
Photo+Auto     & $0.794 \pm 0.036$ & $0.992 \pm 0.045$ & $1.191 \pm 0.054$ & $0.101 \pm 0.005$ & $0.101 \pm 0.005$ & $0.101 \pm 0.005$ \\ \hline
Spec+Cross    & $0.800 \pm 0.003$ & $1.000 \pm 0.003$ & $1.200 \pm 0.003$ & $0.100 \pm 0.001$ & $0.100 \pm 0.001$ & $0.100 \pm 0.001$ \\ \hline
\end{tabular}
\end{table*}

Despite the fact that we are not using any cross-correlations between bins, we can see the parameters for different bins are degenerate. This is because of the effect that \m{\mu} and \m{\sigma} have on the \m{C(l)}. For example, when all the bins are moved in the same direction, the effect is largely to raise or lower the power in each autocorrelation; this can be compensated for by adjusting \m{A_s} and other cosmological parameters appropriately. If however some bins are moved up in redshift, and some down, then the cosmological parameters struggle to compensate for the competing effects. The same is true for \m{\sigma}. This means that these parameters are constrained to move together to some extent. 

We note that the degeneracy between \m{\sigma} and \m{A_s} is being prevented from exercising its full effect because \m{A_s} has been constrained significantly by the CMB information. Nevertheless \m{A_s} is strongly degenerate with \m{\tau_r} (the optical depth at reionisation) and we note that \m{\tau_r} is not well constrained in this instance. Although its effects have been mitigated by its constraint, it is still clear that there is a degeneracy between \m{\sigma} of each bin and all of the cosmological parameters except \m{n_s}. Likewise \m{\mu} is strongly degenerate with \m{\Omega_b}, \m{\Omega_\Lambda}, \m{h}, and \m{w_0}, and is not constrained up to the hard limits of the prior. This means that errors in estimates for binning parameters can propagate into cosmological parameters in a significant way. 

We can understand the degeneracy between the cosmological parameters by considering their effects upon the \m{C(l)}s. The effect of \m{\sigma} is primarily to change the height of the \m{C(l)}s, which creates its degeneracy with \m{A_s}. Likewise, we know that moving \m{\mu} to low redshift boosts power; since \m{\Omega_\Lambda} and \m{h} suppress structure formation, these need to be lowered and \m{\Omega_b} raised to get the C(l)s to match the fiducial model. 





\subsection{Cross correlating with spectroscopic redshifts} \label{cross-spec}

In this section we demonstrate the improvement attainable by cross-correlating with spectroscopic redshift data. Because of the overlap with spectroscopic data, we can show that the properties of the photometric redshift bins are now tightly constrained, and the degeneracies between bins are less pronounced. In most cases, the binning parameters cannot vary widely enough to have a noticeable impact on the cosmological parameters compared to the uncertainty already present. 

We can see that there is increased precision in the cosmological parameters (except for $n_s$, which is almost entirely determined by CMB information here), with most bounds improving by a factor of two or more (Table \ref{cosmo_table}). The slight widening in the posterior distribution for \m{n_s} which can be seen in Figure \ref{Triangle} is most likely due to the additional noise introduced to the galaxy clustering likelihood by looking at larger numbers of bins. Since galaxy number counts do little to constrain \m{n_s} at this level, and bins which are widely separated in redshift may produce correlation functions that are largely noise dominated (since they should be close to zero), this small additional of noise to the likelihood causes some spreading of this parameter. This could be tackled by ignoring widely separated bins if necessary (this would also speed the likelihood calculation by reducing the number of integrations). Binning parameters ($\mu_i, \sigma_i$) have been particularly tightly constrained, allowing us to know their values to percent level or better (Table \ref{bin_table}). Further constraint can be imposed upon them by having more spectroscopic bins to cover a greater fraction of the photometric redshift range, at the cost of computation time. 

\subsection{Constraining higher moments} \label{higher_constraints}

As a benchmark, we also present constraints on the shapes of a single photometric bin from cross correlations with three spectroscopic bins. Here we have a fixed cosmology to simplify the calculations and provide benchmark results for the shape parameters. With cross correlations between more photometric and spectroscopic bins we expect these results to be improved. 

\begin{figure} 
	\includegraphics[width=8cm,keepaspectratio=true]{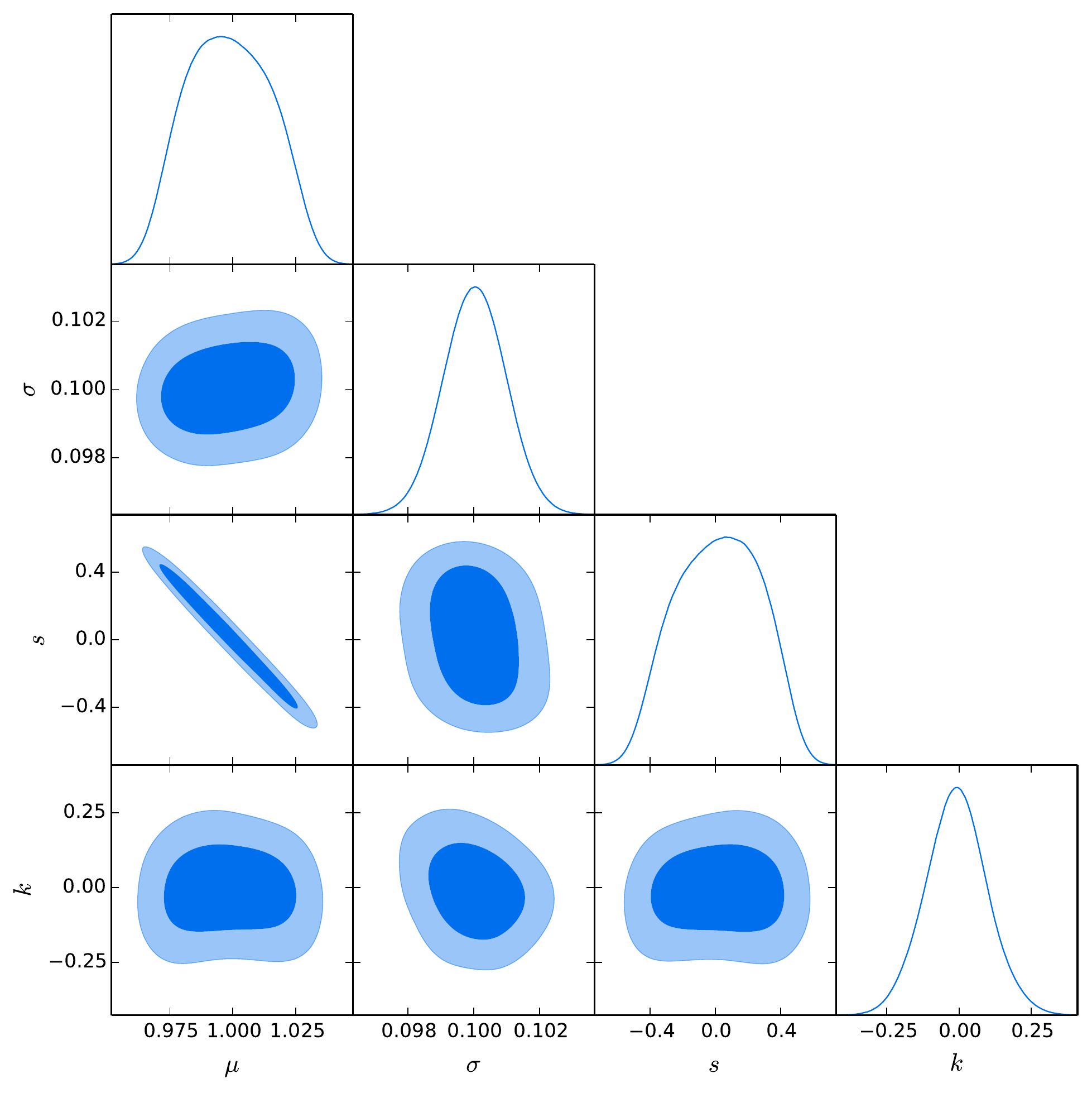}
	\caption{Constraints on the mean, width, and shapes of bins as characterised by $(\mu, \sigma, s, k)$ for a single photometric bin, cross correlated with three spectroscopic bins, with fixed cosmology.}   \label{Bin_Stuff}
\end{figure}

We find that the odd moments are strongly correlated, but we also have less information to constrain the higher moments from the correlation statistics. This may not be a problem unless the higher moments significantly affect the cosmological parameter estimation, although the lack of constraining power of the $C(l)$s suggest that these are not strongly affected by the finer details of the shape of the distributions. The most significant potential problem is the degeneracy between $s$ and $\mu$. This could lead to a spreading of the distribution over $mu$ which may affect the cosmological parameter estimation. In order to combat this one would need to have relatively dense spectroscopic samples throughout the range. In principle, the degeneracy between $\mu$ and $s$ need not be a problem. Since they are strongly degenerate, it is largely the case of the one compensating for the effect of the other. In this case, the cosmological calculation may not be strongly affected even if the uncertainty in $\mu$ increases significantly. We should also bear in mind that the application of the skew transformation alters the mean of the distribution even thought the peak is kept in the same place. This needs to be compensated for by the $\mu$ parameter, and thus the true mean may remain roughly unchanged and be very strongly constrained. We do not expect higher moments to have a very strong impact on cosmological results. 

\section{Conclusions} \label{discussion}

From the results presented in section \ref{results}, we can see that different aspects of the analysis provide distinct benefits. The variation of \m{n(z)}, and its subsequent marginalisation, is essential for the removal of the bias from cosmological inferences. In order utilise the full power of the \m{C(l)} formalism, we must include cross-correlations as well as the well studied auto-correlations; these not only provide us with with much more information (improving our constraining power), but also help us to pin down the relationships between different photometric redshift bins more accurately. Since the photometric redshift parameters display degeneracy with almost all of the cosmological parameters, it is crucial to have these distributions known as well as possible. 

By using a full theoretical likelihood and MCMC approach this study is less idealised than most previous works; nevertheless the assumption of gaussian bins, and the small number of redshift bins, are simplifications that should be addressed in future work. As the analysis is extended with more bins, the computational complexity will increase significantly; the number of cross-correlations to calculate will increase as \m{O(n_{bins}^2)} and the dimensionality of the parameter space as \m{O(n_{bins})}. It is possible to simplify in such cases by only considering cross-correlations between bins which are sufficiently close together (leading to a band diagonal \m{C^{ij}(l)} matrix), since cross-correlations between widely separated bins will contain comparatively little information. Despite the computations intensity, as long as numerical errors remain tamed, we expect that increasing the number of bins and the density of the spectroscopic sample to improve the results. A survey such as Euclid should have a great deal of power to jointly constrain the redshift distribution and the cosmology with minimal disturbance to the confidence intervales for cosmological parameters. 

Errors on redshift binning parameters are now $O(10^{-3})$, even with such sparse spectroscopic data as we have simulated. Errors on the means of photometric bins are at $\pm0.003$ and errors on the width of bins is at $\pm0.001$ (see Table \ref{bin_table}). This is extremely promising for future experiments, providing the possibility to extract reliable and precise cosmological parameters. As we look towards future experiments such as Euclid and LSST, and even with data currently being released from DES, a major focus in cosmology will be the nature of dark energy. The ability to distinguish between a cosmological constant, scalar field theory, modified gravity, or more exotic forms still, will be dependent on having well known redshift distributions, as can be seen by the strong degeneracy between the mean of redshift bins and the parameters \m{\Omega_\Lambda} and \m{w_0}. A bias from improperly calibrated photometric data could easily generate a spurious result. Using this method will help to ensure robust analyses for current and future experiments. 

This method can be applied to any \m{C(l)} signal using photometric redshift bins (such as weak lensing or galaxy clusters) to calibrate their photometric redshift distributions. This means that when applied to future optical surveys, it will be able to benefit much more powerful analyses than the one outlined in this paper, including a larger number of redshift bins, and a combination of signals from different cosmological probes. If the same photometry is used for both number counts and lensing, then both of these cross correlations will contribute to constraining the photometric parameters, as well as constraining the cosmology itself. It will be necessary for future observational work to extend this to non-gaussian distributions, including higher order moments or more generic spline models of \m{n(z)}, in order to model our observed photometric redshifts as best we can. When combined with lensing information, this technique can be applied to achieve improved results in Modified Gravity or Dark Energy studies, where biases can lead to spurious detections and high precision is needed.

\section*{Acknowledgements}

MM is supported by an UCL studentship; FBA acknowledges the Royal Society for a Royal Society University Research Fellowship.

\appendix
\section{Details of n(z) Transformations}\label{transforms}

In this appendix we will describe in detail the transformations made to the \m{n(z)} functions, and the motivations for the heuristic shape manipulation. It is not important that the higher moments are not exactly represented in the same way that the mean and variance of the gaussian are -- these are after all only parameters controlling the shape which will be marginalised over. The important thing is that it can explore a variety of shapes with a small number of parameters. More precise handling of the distributions can be achieved at the cost of increasing the parameter space, which may rapidly make the computation unmanageable without abundant computing resources. 

\subsection{Mean and Variance}

Due to the spline representation used in our computation (see section \ref{codes}) it is simplest to perform a transformation the z axis to a new variable \m{z^\prime = f(z)} for a general (non-Gaussian) distribution. By creating the function \m{n^\prime(z^\prime) = n(z) = n(f^{-1}(z))}, we obtain a distorted distribution in our new variable, which we take to be the new redshift. (This distribution, as all distributions, is normalised before any further calculations are carried out.)

To vary the mean (\m{\mu \rightarrow \mu + \Delta\mu}), we apply the transformation:
\begin{equation}
z^\prime = z + \Delta\mu
\end{equation}
This varies the mean of an arbitrary distribution without affecting any of the higher moments. 
We can also change the standard deviation (\m{\sigma \rightarrow \sigma + \Delta\sigma}) without affecting the mean or higher moments. (The fourth moment \m{m_4} is changed, but not kurtosis \m{\kappa = \frac{m_4}{\sigma^4}}.)
\begin{equation}
z^\prime = \left(1 + \frac{\Delta\sigma}{\sigma}\right)(z-\mu) + \mu
\end{equation}
One is free to change the mean and standard deviation of distributions in whichever order desired as these transformations are commutative. Higher order transforms may also be applied which break symmetries, so care must be taken there as the operations will not be commutative. 

\subsection{Skewness}

To adjust the apparent skewness, we need to stretch the distribution on one side of the mean, and squeeze the distribution on the other. For the sake of simplicity, we write a heuristic skew function controlled by a single parameter \m{s}. (This is to distinguish it from the skewness calculated from the third moment, \m{\gamma}.) The parameter range is defined at \m{-1 < s < 1}. We map from the original redshift coordinate \m{z} to a new coordinate \m{z^\prime} representing the new redshift after the distortion has been taken into account. If the two are identical then we have \m{\frac{dz^\prime}{dz} = 1} everywhere. If we wish to stretch a region then \m{\frac{dz^\prime}{dz} > 1} and to squeeze it we have \m{\frac{dz^\prime}{dz} < 1}. To achieve skewness, we need to smoothly vary from stretched regions on the one side of the mean, to squeezed regions on the other side, with \m{\frac{dz^\prime}{dz}|_{z=\mu} = 1}.

We may choose a simple linear function:
\begin{equation}
 \frac{dz^\prime}{dz} = 1 + \frac{(z-\mu)s}{L} 
\end{equation}

This fulfils the criteria discussed in the range \m{ \mu - L < z < \mu + L }. After these points we fix \m{\frac{dz^\prime}{dz}|_{z < \mu-L} = \frac{dz^\prime}{dz}|_{z = \mu-L} } and \m{\frac{dz^\prime}{dz}|_{z > \mu+L} = \frac{dz^\prime}{dz}|_{z = \mu+L} }. This is because otherwise we rapidly end up with very extreme stretching or squeezing of the distribution. Here we have an extra free parameter, \m{L} (the lengthscale of the skewness function). To avoid overburdening the routine with extra parameters we generically set this to \m{L = \frac{3\sigma}{4}}. The resulting function is found by integrating these expressions with the condition that \m{\mu^\prime = \mu}. (This means that skewness does not interfere with the peak of the distribution, but will change the mean; it may also interfere with standard deviation. Standard deviation and mean can separately be readjusted to remove this degeneracy if desired.)

\begin{equation}
z^\prime = \begin{cases} 
			\frac{s}{L} \left( \frac{1}{2} (z + \mu)^2 - \mu \right) + z, & |z-\mu| \le L \\
			(\mu+L)^\prime + (1+s)(z-(\mu+L)), & z-\mu > L \\
			(\mu -L)^\prime + (1-s)(z - (\mu-L)), & z - \mu < -L
		\end{cases}
\end{equation}

\begin{figure} 
	\includegraphics[width=8cm,keepaspectratio=true]{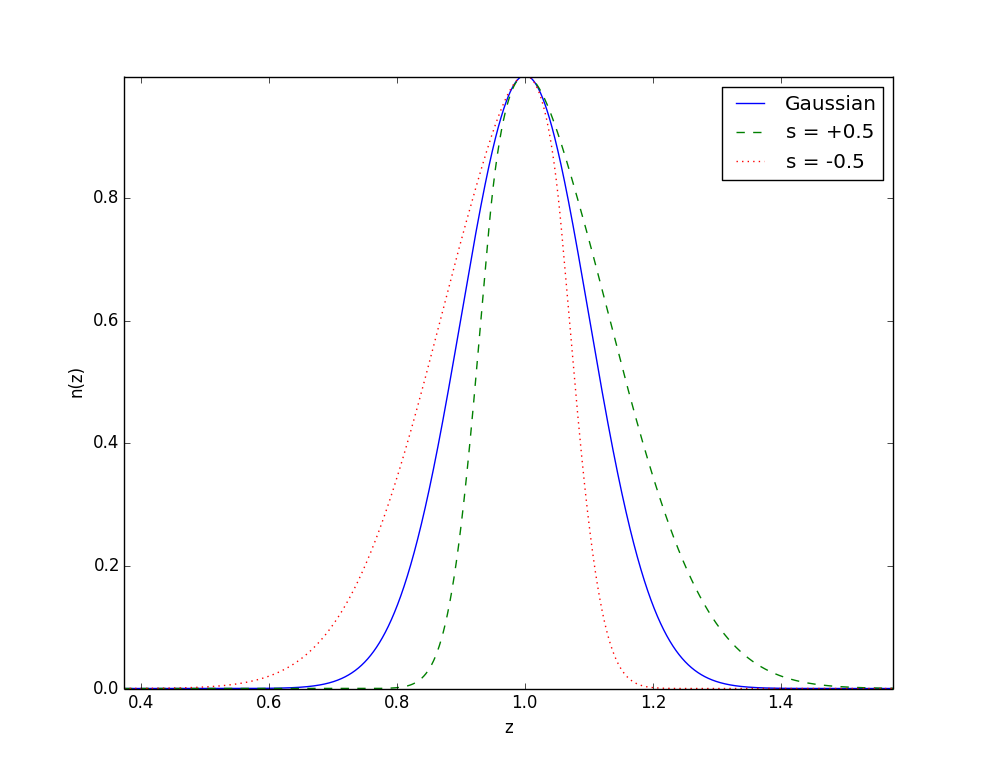}
	\caption{Results of applying positive and negative skew transformations to a gaussian photometric bin.}
		\label{skews}
\end{figure}

\subsection{Kurtosis}

Kurtosis is handled in a similar way to skewness, by a heuristic function controlled by a single parameter \m{k} (different to the exact kurtosis, traditionally \m{\kappa}) which varies \m{-1 < k < 1}. This case is symmetric, and we wish to stretch the distribution close to the mean, and squeeze it further away from the mean (or vice versa). 

In this case we again need to choose length scales. We choose to have the transition from stretched to squeezed regions (i.e. \m{\frac{dz^\prime}{dz} = 1}) at \m{(z-\mu) = \sigma}, and then fix the transformations (as with the skewness above) at \m{(z-\mu) = 2\sigma}. This requires in the below expression \m{L = \sigma}. 

For simplicitly we again choose linear relations. 
\begin{equation}
 \frac{dz^\prime}{dz} = (1+k) - \frac{|z-\mu|}{L}k
\end{equation}

We then integrate as before, choosing \m{\mu^\prime = \mu}. Once again, the standard deviation can be separately adjusted for if desired. 
\begin{equation}
z^\prime = \begin{cases} 
			z + k(z-\mu) - k\frac{(z-\mu)^2}{2\sigma}, & \mu \le z \le \mu+2\sigma \\
			z + k(z-\mu) + k\frac{(z-\mu)^2}{2\sigma}, & \mu-2\sigma \le z \le \mu \\
			\mu + \sigma + (1-k)(z-\mu-2\sigma), & z \ge \mu+2\sigma \\
			\mu - \sigma + (1-k)(z-\mu+2\sigma), & z \le \mu-2\sigma
		\end{cases}
\end{equation}

\begin{figure} 
	\includegraphics[width=8cm,keepaspectratio=true]{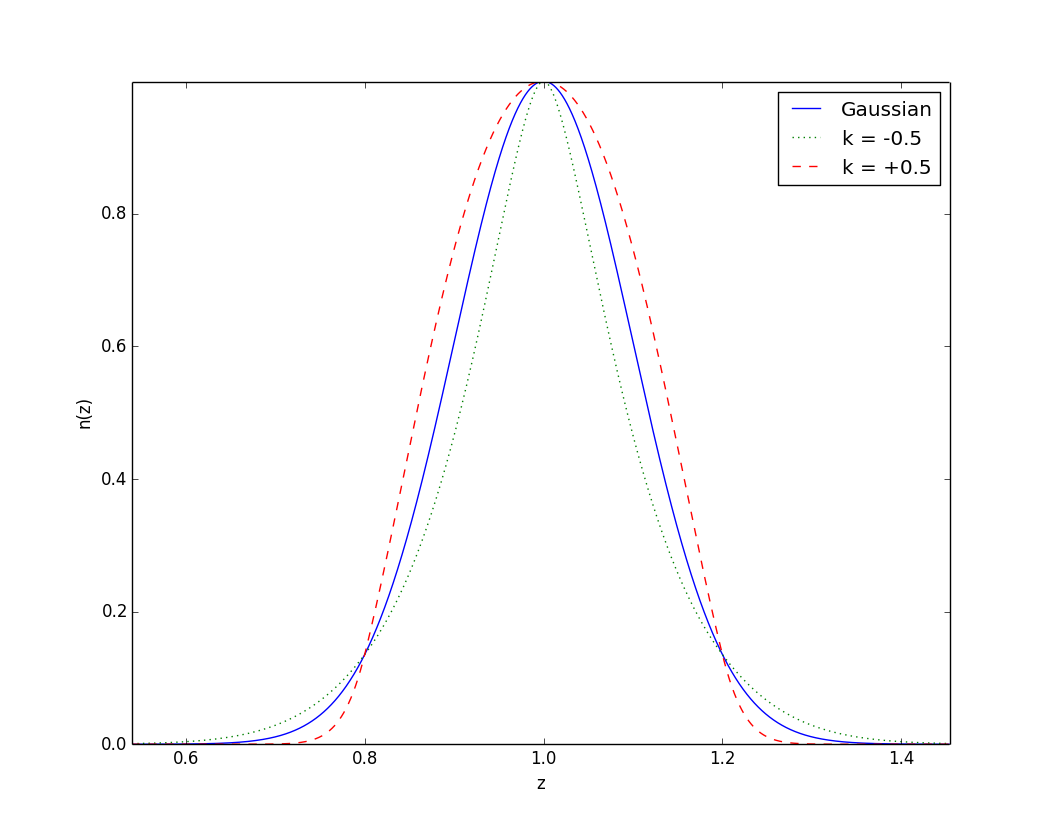}
	\caption{Results of applying positive and negative kurtosis transformations to a gaussian photometric bin.}
		\label{skews}
\end{figure}





\begin{thebibliography}{99}

\bibitem[\protect\citeauthoryear{Abdalla \& Rawlings}{2008}]{Fil_Prob}
Abdalla F., Rawlings S., 2008, MNRAS, 381, 1313

\bibitem[\protect\citeauthoryear{Abdalla et al.}{2008}]{photoz_compare}
Abdalla F. B., Banerji M., Lahav O., Rashkov V., 2008, MNRAS, 417, 1891

\bibitem[\protect\citeauthoryear{Banerji et al.}{2008}]{ANN_redshifts}
Banerji M., Abdalla F., Lahav O., Lin H., 2008, MNRAS, 386, 1219

\bibitem[\protect\citeauthoryear{Benjamin et al.}{2010}]{Benjamin_Contamination}
Benjamin J. et al, 2010, MNRAS, 408, 1168

\bibitem[\protect\citeauthoryear{Blake et al.}{2004}]{Blake_1}
Blake C., Ferreira P. G., Borrill J., 2004, MNRAS, 351, 923

\bibitem[\protect\citeauthoryear{Blake et al.}{2007}]{Blake_2}
Blake C. Collister A., Bridle S., Lahav O., 2007, MNRAS, 374, 1527

\bibitem[\protect\citeauthoryear{Blas et al.}{2011}]{CLASS}
Blas D., Lesgourgues J., Tram T., 2011, JCAP, 07, 034

\bibitem[\protect\citeauthoryear{Bucher et al.}{2002}]{ProbRef}
Bucher M., 2002, Phys. Rev. D 66, 023528

\bibitem[\protect\citeauthoryear{Clerkin et al.}{2014}]{Clerkin_bias}
Clerkin L. et al., 2014, MNRAS 448, 1389

\bibitem[\protect\citeauthoryear{Huterer et al.}{2001}]{Huterer_Cl}
Huterer D., 2001, ApJ, 555, 547

\bibitem[\protect\citeauthoryear{Huterer et al.}{2004}]{Huterer_needs}
Huterer D., et al., 2004, ApJ, 615, 595

\bibitem[\protect\citeauthoryear{Kirk et al.}{2013}]{CrossCorr}
Kirk D., Lahav O. et al., 2013, MNRAS, 438, 2218

\bibitem[\protect\citeauthoryear{Lahav et al.}{2009}]{DESForecast}
Lahav O., Kiakotou A., Abdalla F. B., Blake C., 2009, MNRAS, 405, 168

\bibitem[\protect\citeauthoryear{Matthews \& Newman}{1982}]{Matthews_Reconstruction}
Matthews D., Newman J., 2010, ApJ, 721, 456

\bibitem[\protect\citeauthoryear{McQuinn \& White}{2013}]{McQuinn}
McQuinn M., White M., 2013, MNRAS, 433, 2857

\bibitem[\protect\citeauthoryear{Menard et al.}{2014}]{Menard_Calibration}
Menard B. et al., 2011, arXiv:1303.4722v2

\bibitem[\protect\citeauthoryear{Newman}{2008}]{Newman_CC}
Newman J., 2008, arXiv:0805.1409

\bibitem[\protect\citeauthoryear{Newman et al.}{2012}]{SpecRef}
Newman J. A. et al., 2012, Astroparticle Physics, 63, 81

\bibitem[\protect\citeauthoryear{Peebles}{1973}]{Peebles}
Peebles P. J. E., 1973, ApJ, 185, 413

\bibitem[\protect\citeauthoryear{Rhodes et al.}{2014}]{Rhodes_CC}
Rhodes J. et al., 2014,  arXiv:1309.5388v3

\bibitem[\protect\citeauthoryear{Sadeh et al.}{2015}]{ANNz2}
Sadeh I., Abdalla F., Lahav O., 2015, arXiv:1507.00490v1

\bibitem[\protect\citeauthoryear{Schmidt}{2006}]{Schmidt_Recovery}
Schmidt S. et al., 2013, MNRAS, 431, 3307

\bibitem[\protect\citeauthoryear{Schulz}{2015}]{Schulz_Calibration}
Schulz A.E., 2006, ApJ, 724, 1305

\bibitem[\protect\citeauthoryear{SDSS Collaboration}{2008}]{BOSS}
SDSS Collaboration, 2008, Astron. J, 142, 72

\bibitem[\protect\citeauthoryear{Thomas et al.}{2012}]{Thomas}
Thomas S. A., Abdalla F., Lahav O., 2012, MNRAS, 412, 1169

\bibitem[\protect\citeauthoryear{Zheng \& Zhang}{2012}]{photoz_review}
Zheng H. \& Zhang Y., 2012, Proc. SPIE 8451, Software and Cyberinfrastructure for Astronomy II, 845134

\end{thebibliography}





\bsp	
\label{lastpage}
\end{document}